\documentclass[12pt]{article}
\usepackage{graphicx}
\usepackage{epstopdf}
\usepackage{amsmath}
\usepackage{amsfonts}
\usepackage{amssymb}
\usepackage{color}
\usepackage{mathrsfs}

\newcommand{\be}{\begin{equation}}
\newcommand{\ee}{\end{equation}}
\newcommand{\barr}{\begin{array}}
\newcommand{\earr}{\end{array}}

\usepackage{latexsym}
\usepackage{amsthm}
\usepackage{amsmath}
\usepackage{graphicx}
\usepackage{amssymb}
\usepackage{bbm}
\newcommand{\gsim}{\lower.7ex\hbox{$\;\stackrel{\textstyle>}{\sim}\;$}}
\newcommand{\lsim}{\lower.7ex\hbox{$\;\stackrel{\textstyle<}{\sim}\;$}}

\newcommand{\bea}{\begin{eqnarray}}
\newcommand{\eea}{\end{eqnarray}}

\newcommand{\comment}[1]{}

\def\e{\mathrm{e}}

\def\half{{1\over 2}}

\def\half{{1\over 2}}
\def\({\left(}
\def\){\right)}
\def\[{\left[}
\def\]{\right]}
\def\e{\begin{equation}}
\def\q{\end{equation}}
\def\m{\begin{eqnarray}}
\def\n{\end{eqnarray}}

\begin{document}

\setcounter{page}{1} \baselineskip=15.5pt \thispagestyle{empty}
\vfil

\begin{center}

{\Large \bf An accurate determination of the Hubble constant \\
\vspace{3mm}
from Baryon Acoustic Oscillation datasets}
\\[0.7cm]
{Cheng Cheng$^{1,2}$ and Qing-Guo Huang$^1$}
\\[0.7cm]

{\normalsize { \sl $^{1}$ State Key Laboratory of Theoretical Physics, Institute of Theoretical Physics, \\ Chinese Academy of Science, Beijing 100190, China}}\\
\vspace{.2cm}

{\normalsize { \sl $^{2}$  University of the Chinese Academy of Sciences, Beijing 100190, China}}
\vspace{.3cm}

\end{center}

\vspace{.8cm}

\hrule \vspace{0.3cm}
{\small  \noindent \textbf{Abstract} \\[0.3cm]
Even though the Hubble constant cannot be significantly determined by the low-redshift Baryon Acoustic Oscillation (BAO) data alone, it can be tightly constrained once the high-redshift BAO data are combined. Combining BAO data from 6dFGS, BOSS DR11 clustering of galaxies, WiggleZ and $z=2.34$ from BOSS DR11 quasar Lyman-$\alpha$ forest lines, we get $H_0=68.17^{+1.55}_{-1.56}$ km s$^{-1}$ Mpc$^{-1}$. In addition, adopting the the simultaneous measurements of $H(z)$ and $D_A(z)$ from the two-dimensional two-point correlation function from BOSS DR9 CMASS sample and two-dimensional matter power spectrum from SDSS DR7 sample, we obtain $H_0=68.11\pm1.69$ km s$^{-1}$ Mpc$^{-1}$. Finally, combining all of the BAO datasets, we conclude $H_0=68.11\pm 0.86$ km s$^{-1}$ Mpc$^{-1}$, a $1.3\%$ determination.
}
 \vspace{0.3cm}
\hrule

\vspace{8cm}

\newpage
\section{Introduction}

\begin{figure}[!htb]
\begin{center}
\includegraphics[width=12 cm]{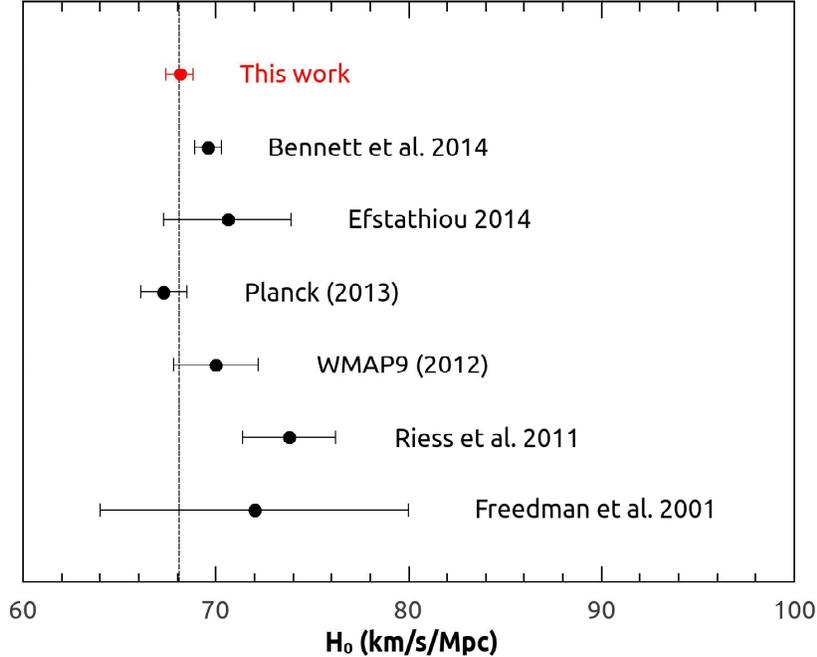}
\end{center}
\caption{Comparison of different $H_0$ measurements.}
\label{fig:h0s}
\end{figure}

Since Edwin Hubble firstly published the linear correlation between the apparent distances to galaxies and their recessional redshift in $1929$ \cite{Hubble:1929ig}, the measurement of Hubble constant $H_0$  became a central goal in cosmology. $H_0$ measures the present expansion rate of the universe and is closely related to the components of the universe. Moreover, the inverse of $H_0$ roughly sets the size and age of the universe. Its value was estimated between $50$ and $100$ km s$^{-1}$ Mpc$^{-1}$ for decades, until Hubble Space Telescope (HST) and its Key project released their results in \cite{HST_Freedman_2000} which was the first time that  $H_0$ was measured accurately, namely $H_0 = 72 \pm 8$ km s$^{-1}$ Mpc$^{-1}$, a roughly $11\%$ determination. 
This result was significantly improved by Riess et al. in 2011 \cite{HST_Riess_2011} where $H_0 = 73.8 \pm 2.4$ km s$^{-1}$ Mpc$^{-1}$, a $3\%$ determination. 
On the other hand, the Hubble constant can be determined by the cosmic microwave background (CMB) data indirectly. Assuming a flat universe, the nine-year Wilkinson Microwave Anisotropy Probe (WMAP9) data alone \cite{wmap9} give a $3\%$ determination, i.e. $H_0 = 70.0 \pm 2.2$ km s$^{-1}$ Mpc$^{-1}$, in the $\Lambda$CDM model in the end of 2012. In the early of 2013, Planck \cite{planck} released its first result and found the derived Hubble constant of $67.3 \pm 1.2 $ km s$^{-1}$ Mpc$^{-1}$ which is roughly $2.5\sigma$ tension with the Riess et al. cosmic distance ladder measurement \cite{HST_Riess_2011} at low redshift. 
In order to clarify such a tension, Efstathiou re-analyzed the Riess et al. \cite{HST_Riess_2011} Cepheid data in \cite{H0_Efstathiou}. Based on the revised geometric maser distance to NGC 4258, he found $H_0 = 70.6 \pm 3.3$ km s$^{-1}$ Mpc$^{-1}$ which is consistent with both HST \cite{HST_Riess_2011} and Planck \cite{planck}. 
Recently combining Baryon Acoustic Oscillation (BAO) results of 6dF Galaxy Surver (6dFGS) \cite{6df}, Baryon Oscillation Spectroscopic Survey (BOSS) Data Release 11 (DR11) \cite{boss_dr11,bao_234}, distance ladder $H_0$ determination \cite{HST_Riess_2011}, WMAP9 \cite{wmap9} and supplementary CMB data at small angular scales from two ground-based experiments (Atacama Cosmology Telescope and South Pole Telescope), Bennett et al. \cite{H0_Bennett} got the a more accurate determination of $H_0=69.6\pm0.7$ km s$^{-1}$ Mpc$^{-1}$. All of these results are summarized in Fig.~\ref{fig:h0s}. 
Even though the different measurements give roughly the same value of $H_0$, there are still $1\sim 2 \sigma$ discrepancies. So it is still worthy determining the Hubble constant from other independent cosmological experiments. In this paper we adopt the BAO data alone to constrain the Hubble constant, and find $H_0=68.11\pm0.86$ km s$^{-1}$ Mpc$^{-1}$. See the red point in Fig.~\ref{fig:h0s}.

Measurement of BAO is a very important tool for probing cosmology. 
Before recombination and decoupling the universe consisted of a hot plasma of photons and baryons tightly coupled via Thomson scattering. The competition of radiation pressure and gravity sets up oscillations in the photon fluid. At recombination the universe became neutral and the pressure on the baryons disappeared, and this abrupt change imparted a slight over-density of baryons on the length scale given by the distance that the sound waves could have traveled since the big bang, i.e. the sound horizon $r_s(z_d)$. Since the baryons and dark matter interact though gravity, the dark matter also preferentially clumps on this scale. Therefore, BAO imparts a characteristic signal in the matter power spectrum on the scale of the sound horizon at recombination. This signal in the matter power spectrum can be used as a ``standard ruler" to map out the evolution of the Hubble parameter $H(z)$ and the angular diameter distance $D_A(z)$ at redshift $z$. 
Usually the Hubble parameter and the angular diameter distance cannot be extracted simultaneously from the BAO data. 
In \cite{Eisenstein:2005su} an effective distance $D_V\propto (D_A(z)^2H^{-1}(z))^{1/3}$ was introduced according to the different dilation scales for $H(z)$ and $D_A(z)$. 
Unfortunately the low-redshift BAO data, for example $D_V(z)/r_s(z_d)$, are insensitive to the Hubble constant, and hence the low-redshift BAO data can tightly constrain the matter density parameter $\Omega_m$, but not the Hubble constant, by themselves. But this degeneracy of the Hubble constant can be significantly broken once the high-redshift BAO data are taken into account.

On the other hand, in principle the Hubble parameter $H(z)$ and the angular diameter distance $D_A(z)$ can be extracted simultaneously from the data through the measurement of the BAO scale in the radial and transverse directions. In \cite{Chuang:2011fy,Chuang:2012ad,Chuang:2012qt} Chuang and Wang made significant improvements in modeling for the two-dimensional two-point correlation function (2d2pCF) of galaxies, and succeeded in simultaneously measuring $H(z)$ and $D_A(z)$ from data without assuming a dark energy or a flat Universe. Recently the similar method and model have been applied to measure $H(z)$, $D_A(z)$ and the physical matter density $\Omega_m h^2$ from the anisotropic galaxy clustering of DR9 CMASS sample of the SDSS-III BOSS at the effective redshift $z=0.57$ \cite{Chuang:2013hya}. In addition, a method to measure $H(z)$ and $D_A(z)$ simultaneously from the two-dimensional matter power spectrum (2dMPS) was proposed by Hemantha et al. in \cite{Hemantha:2013sea}. Applying this method to Sloan Digital Sky Survey (SDSS) DR7, Hemantha et al. simultaneously constrained $H(z)$, $D_A(z)$ and $\Omega_m h^2$ as well. These two measurements can provide a significant constraint on the cosmological parameters, e.g. the Hubble constant, themselves.


In this paper we assume a spatially flat universe containing a cosmological constant and cold dark matter, namely a concordance $\Lambda$CDM cosmological model. We will introduce the BAO data and explain our methodology in Sec.~2. Our main results will be presented in Sec.~3. Summary and discussion will be given in Sec.~4.

\section{BAO data and Methodology} 

BAO provides an independent way to determine cosmological parameters. 
The BAO signal is a standard ruler such that the length of the sound horizon can be measured as a function of redshift. This measures two cosmological distances: $D_A(z)/r_s(z_d)$ (the correlations of two spatial dimensions orthogonal to the direction of sight) and $H(z)r_s(z_d)$ (the fluctuation of one dimension along the direction of sight). 

In a spatially flat universe the angular diameter distance is given by 
\e
\label{da}
D_A(z) = \frac {1} {1+z} \int_0^z \frac {dz'} {H(z')}, 
\q
where $H(z)$ is related to the Hubble constant $H_0$ by 
\e
\frac {H(z)}{H_0} = \[ \Omega_{r}(1+z)^4+\Omega_m(1+z)^3+(1-\Omega_{r}-\Omega_m) f(z) \]^\half, 
\q
here $f(z)\equiv {\rho_{\rm de}(z)}/{\rho_{\rm de}(0)}$ depends on dark energy model, i.e. 
\e
f(z)= (1+z)^{3(1+w_0+w_a)}\exp \[{- \frac{3w_az} {1+z}}\], 
\q
in the $w_0w_a$CDM model \cite{cpl} where $w=p_{\rm de}/\rho_{\rm de}$ is the equation of state parameter of dark energy which is parameterized by $ w(z) = w_0 + w_a z/(1+z)$. For the $\Lambda$CDM model, $w_0=-1$ and $w_a=0$. 
In this paper we adopt Eisenstein \& Hu  \cite{Hu&E} form to calculate the sound horizon at redshift $z_d$ which is the time when baryons decoupled from the Compton drag of photons, namely 
\e
\label{rs}
r_s(z_d) = \frac {1} {\sqrt 3} \int_0^{\frac 1 {1+z_d}} \frac {da}  {a^2H(a) \sqrt{1+ \frac {3 \Omega_b} {4\Omega_{\gamma} }a}}, 
\q
where 
\e
z_d = \frac {1291(\Omega_m h^2)^{0.251}}{1+0.659(\Omega_m h^2)^{0.828}}[1+b_1(\Omega_b h^2)^{b_2}], 
\q
and 
\m
b_1 &=& 0.313(\Omega_m h^2)^{-0.419}[1+0.607(\Omega_m h^2)^{0.674}], \\
b_2 &=& 0.238(\Omega_m h^2)^{0.223}. 
\n
Since the Hubble parameter changes different from the angular diameter distance, the dilation scale is usually treated as the cube root of the product of the radial dilation times the square of the transverse dilation \cite{Eisenstein:2005su}, namely 
\e
\label{dv}
D_V(z) \equiv \left [(1+z)^2D_A^2(z)\frac {z} {H(z)}\right]^{\frac 1 3}, 
\q 
which is the so-called volume-averaged effective distance.

In this paper, $\Omega_b h^2$ and $\Omega_{\gamma} h^2$ are fixed as their best fit values of Planck, namely 
\m
\Omega_b h^2 &=& 0.02203 \\
\Omega_{\gamma} h^2 &=&  2.46 \times 10^{-5}.
\n
The present energy density of radiations is related to $\Omega_\gamma$ by 
\e
\Omega_{r} = \Omega_{\gamma}  (1+0.2271N_{\rm eff}) = 4.16 \times 10^{-5} h^{-2}, 
\q
where $N_{\rm eff}=3.046$ is the effective number of neutrinos in standard model. The changes of $\Omega_b h^2$ within its error bars do not substantially shift our results.


\subsection{BAO-I}

In \cite{6df} the distance-redshift relation at the effective redshift $z_{\rm eff}=0.106$ is $r_s(z_d)/D_V=0.336\pm 0.015$ from the large-scale correlation function of the 6dF Galaxy Survey (6dFGS) BAO.

Fitting for the position of the acoustic features in the correlation function and matter power spectrum of BAO in the clustering of galaxies from BOSS DR11, Anderson et al. get  $D_V(z=0.32)(r_s(z_d)_{\rm fid}/r_s(z_d)) = 1264\pm25$ Mpc and $D_V(z=0.57)(r_s(z_d)_{\rm fid}/r_s(z_d)) = 2056\pm20$ Mpc in \cite{boss_dr11}. Here $r_s(z_d)_{\rm fid} = 153.19$ Mpc \footnote{In this paper, we use Eisenstein \& Hu form. So we adopt $r_s(z_d)_{\rm fid} = 153.19$ Mpc. } in their fiducial cosmology. When fitting cosmological models, we adopt $D_A(z=0.57)(r_s(z_d)_{\rm fid}/r_s(z_d))=1421\pm 20$ Mpc and $H(z=0.57)(r_s(z_d)/r_s(z_d)_{\rm fid})=96.8\pm 3.4$ km s$^{-1}$ Mpc$^{-1}$, with a correlation coefficient between $D_A$ and $H$ of $0.539$, which is recommended by Anderson et al. in \cite{boss_dr11}.  

Reconstructing the baryonic acoustic feature from the WiggleZ Dark Energy Survey, the model independent distances $D_V(r_s(z_d)_{\rm fid}/r_s(z_d))$ given by Kazin et al. in \cite{wigglez} are $1716\pm83$ Mpc, $2221\pm101$ Mpc, $2516\pm86$ Mpc at effective redshifts $z=0.44,\ 0.6,\ 0.73$ respectively. The fiducial cosmology adopted in \cite{wigglez} implies $r_s(z_d)_{\rm fid} = 148.6$ Mpc.

From BOSS DR11 latest released sample, Delubac et al. figure out the BAO feature in the flux-correlation function of the Lyman-$\alpha$ forest of high-redshift quasars in \cite{bao_234}, and find $\alpha_{\|} = 1.054^{+0.032}_{-0.031} $ and $ \alpha_{\perp} = 0.973^{+0.056}_{-0.051} $ at the effective redshift $z=2.34$, where 
\m
\alpha_{\|}&=& {c/(H(z)r_s(z_d))\over c/(H(z)r_s(z_d))_{\rm fid}}, \\
\alpha_{\perp} &=& {D_A(z)/r_s(z_d) \over (D_A(z)/r_s(z_d))_{\rm fid}}. 
\n
In \cite{bao_234} Delubac et al. recommend using $\alpha_{\|}^{0.7} \alpha_{\perp}^{0.3} = 1.025\pm0.021$, which is the most precisely determined combination, to fit the cosmological models. According to their fiducial cosmology, $c/(H(z=2.34)r_s(z_d))_{\rm fid} = 8.708$ and $(D_A(z=2.34)/r_s(z_d))_{\rm fid} = 11.59$.

\begin{figure}[!htb]
\begin{center}
\includegraphics[width=12 cm]{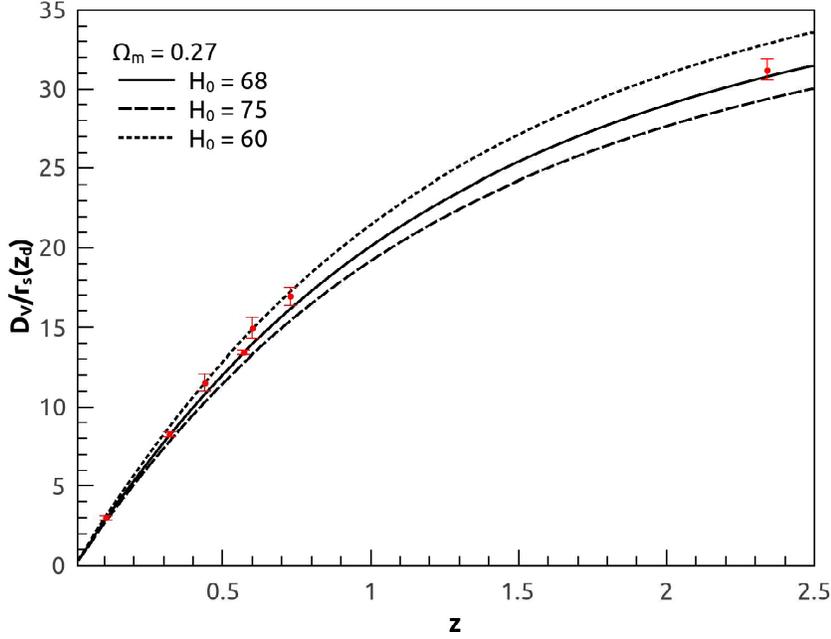}
\end{center}
\caption{$D_V(z)/r_s(z_d)$ varies with $z$. Here $\Omega_m$ is fixed as $0.27$. The solid, dashed and dotted lines correspond to $H_0 = 68, 75, 60$ km/s/Mpc respectively. It implies that the low-redshift BAO alone cannot determine $H_0$ precisely, but $H_0$ can be tightly constrained once the high-redshift BAO data are combined.}
\label{fig:dvtors}
\end{figure}

The distance ratio $D_V(z)/r_s(z_d)$ for the BAO-I data are accumulated in Table.~\ref{table_baoi} and Fig.~\ref{fig:dvtors}. 
\begin{table*}[!hts]
\centering
\renewcommand{\arraystretch}{1.5}
\begin{tabular}{cccc}
\hline\hline
 z & $D_V(z)/r_s(z_d)$ & experiment & reference \\
\hline
0.106 & $2.976\pm 0.133$ & 6dFGS & \cite{6df} \\
0.32 & $8.251\pm 0.163$ &BOSS DR11 & \cite{boss_dr11} \\
0.57 & $13.421\pm 0.131$ &BOSS DR11 & \cite{boss_dr11} \\
0.44 & $11.548 \pm 0.559$ & WiggleZ&\cite{wigglez}\\
0.6 &$14.946 \pm 0.680$ &WiggleZ &\cite{wigglez} \\
0.73&$16.931\pm 0.579$ &WiggleZ & \cite{wigglez} \\
2.34 &$31.233\pm 0.663$ &BOSS DR11&\cite{bao_234}\\
\hline
\end{tabular}
\caption{The distance ratio $D_V(z)/r_s(z_d)$ for the BAO-I datasets. }
\label{table_baoi}
\end{table*}
Keeping $\Omega_m=0.27$ fixed, we plot $D_V(z)/r_s(z_d)$ versus redshift $z$ in Fig.~\ref{fig:dvtors} where the solid, dashed and dotted lines represent $H_0 = 68, 75, 60$ km/s/Mpc respectively. It shows that these three curves trend to converge at low redshift. That is why the low-redshift BAO data alone can not be used to constrain the Hubble constant. But these three curves diverge at high redshift. So the combination of the low and high-redshift BAO data can be used to precisely determine the Hubble constant $H_0$.

\subsection{BAO-II}

In \cite{Chuang:2013hya} Chuang et al. analyzed the broad range shape of the monopole and quadrupole 2d2pCF from BOSS DR9 CMASS sample, and obtained the constraints at the effective redshift $z=0.57$: $\{H(0.57),\ D_A(0.57),\ \Omega_m h^2 \} = \{ 87.6^{+6.7}_{-6.8}\ \hbox{km/s/Mpc},\ 1396\pm 73\ \hbox{Mpc},\ 0.126^{+0.008}_{-0.010}\}$ and their covariance matrix 
\e 
\( 
   \begin{array}{ccc} 
     $0.0385~~~~~~~-0.001141~~~-13.53$\\
     $-0.001141~~0.0008662~~~3.354$\\
     $-13.53~~~~~~~3.354~~~~~~~~ 19370$\\
   \end{array}
 \). 
\q

In \cite{Hemantha:2013sea} Hemantha et al. presented a method to measure $H(z)$ and $D_A(z)$ simultaneously from the 2dMPS from galaxy surveys with broad sky coverage. Adopting SDSS DR7 sample, they obtained the measurements of 
$\{H(0.35),\ D_A(0.35),\ \Omega_m h^2 \} = \{ 81.3\pm 3.8\ \hbox{km/s/Mpc},\ 1037 \pm 44\ \hbox{Mpc},\ 0.1268\pm0.0085 \}$ with covariance matrix 
\e
\( 
   \begin{array}{lll} 
$0.00007225~~~~-0.169609~~~~0.01594328$\\
$-0.169609~~~~~~1936~~~~~~~~~~67.03048$\\
$0.01594328~~~~67.03048~~~~~14.44$\\
   \end{array}
 \). 
\q

\section{Analysis}

\subsection{Consistency of BAO data}

\begin{figure}[!htb]
\begin{center}
\includegraphics[width=16 cm]{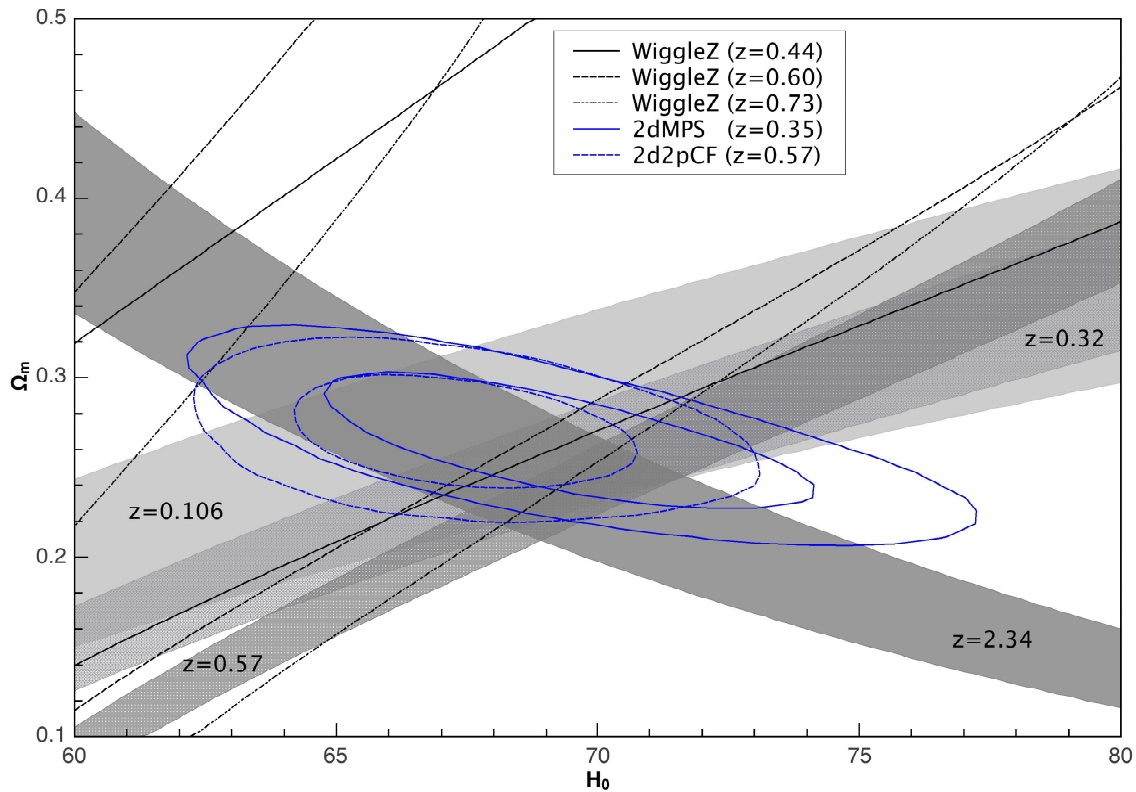}
\end{center}
\caption{BAO $1\sigma$ constraints for redshifts 0.106 from 6dFGS \cite{6df}, 0.32 and 0.57 from BOSS DR11 clustering of galaxies \cite{boss_dr11}, WiggleZ $(z=0.44,\ 0.60,\ 0.73)$ \cite{wigglez}, and $z=2.34$ $(\alpha_{\|}^{0.7} \alpha_{\perp}^{0.3})$ from BOSS DR11 quasar Lyman-$\alpha$ forest lines \cite{bao_234}. The solid and dashed blue contours correspond to the constraints from SDSS DR7 2dMPS at the effective redshift $z=0.35$ \cite{Hemantha:2013sea} and BOSS DR9 2d2pCF at $z=0.57$ \cite{Chuang:2013hya} respectively. }
\label{fig:consistency}
\end{figure}

Since the BAO data mostly constrain the expansion history of the universe which is determined by two parameters ($\Omega_m$ and $H_0$) in the concordance $\Lambda$CDM cosmology, we explore the constraints on these two parameters explicitly.  
In order to check the consistency of different BAO data, the representative BAO measurements at different redshifts from both BAO-I and BAO-II datasets are plotted in Fig.~\ref{fig:consistency} individually. 

From Fig.~\ref{fig:consistency}, one can clearly see that even though different BAO measurements at different redshifts have different degeneracy directions, there is no significant tension among them. 

\subsection{Constraints on the Hubble constant from different BAO datasets}

\begin{figure}[!htb]
\begin{center}
\includegraphics[width=12 cm]{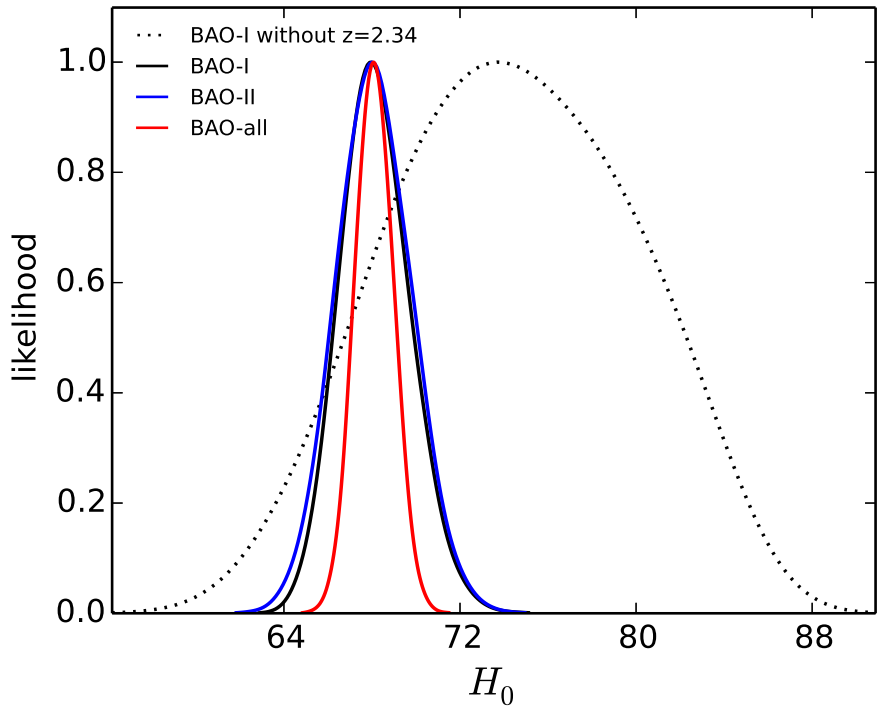}
\end{center}
\caption{The likelihoods of $H_0$ from different BAO datasets. The black dotted and solid curves correspond to the constraints from BAO-I without $z=2.34$ and all BAO-I data respectively. The blue one is the constraint from BAO-II. The likelihood of $H_0$ from all of BAO data is illustrated by the red curve. }
\label{fig:h0}
\end{figure}

The constraints on $H_0$ from different BAO datasets are illustrated in Fig.~\ref{fig:h0}. 

According to the previous arguments, the value of $H_0$ cannot be significantly constrained by the low-redshift BAO-I data only. Adopting the BAO-I dataset without $z=2.34$, $H_0=74.32^{+5.87}_{-5.73}$ km s$^{-1}$ Mpc$^{-1}$. See the black dashed curve in Fig.~\ref{fig:h0}. 
But it can be tightly constrained once one high-redshift BAO (z=2.34) is added. See the black solid curve in Fig.~\ref{fig:h0}. From all of the BAO-I data, the constraint on the Hubble constant is $H_0=68.17^{+1.55}_{-1.56}$ km s$^{-1}$ Mpc$^{-1}$, a $2.3\%$ determination. 

Since the Hubble parameter $H(z)$ and the angular diameter distance $D_A$ can be measured simultaneously from the two-dimensional two-point correlation function and two-point matter power spectrum, the Hubble constant is precisely determined by the two BAO-II datasets alone, namely $H_0=68.11\pm1.69$ km s$^{-1}$ Mpc$^{-1}$, a $2.5\%$ determination, which is roughly the same as that from BAO-I. 

Since BAO-I and BAO-II are consistent with each other, combining both of them, we reach a $1.3\%$ determination of the Hubble constant, namely $H_0=68.11\pm 0.86$ km s$^{-1}$ Mpc$^{-1}$.

\begin{figure}[!htb]
\begin{center}
\includegraphics[width=12 cm]{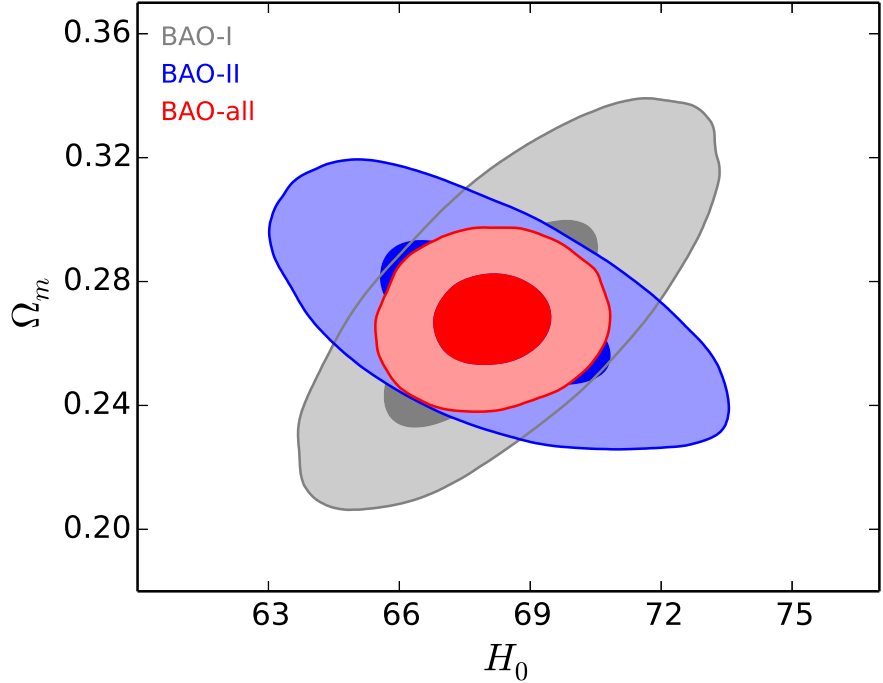}
\end{center}
\caption{The contour plot of $H_0$ vs. $\Omega_m$ from BAO-I, BAO-II and BAO-all respectively. }
\label{fig:bao}
\end{figure}

The contour plots of $H_0$ vs. $\Omega_m$ from BAO-I, BAO-II and BAO-all show up in Fig.~\ref{fig:bao}. 
It is interesting that the contour plots from BAO-I and BAO-II seems orthogonal to one another roughly. Therefore it is quite effective to constrain the cosmological parameters, e.g. $H_0$ and $\Omega_m$, once both BAO-I and BAO-II are combined. 
The numerical results are summarized in Table \ref{tab:bao_all}. 
\begin{table}[!hts]
\centering
\renewcommand{\arraystretch}{1.5}
\begin{tabular}{c|c|c}
 \hline\hline
& $H_0$ (km/s/Mpc) & $\Omega_m$\\
\hline
BAO-I without $z=2.34$ & $74.32^{+5.87}_{-5.73}$ & $0.351^{+0.081}_{-0.077}$ \\
BAO-I & $68.17^{+1.55}_{-1.56}$ &  $0.268^{+0.022}_{-0.021}$ \\
BAO-II  & $68.11\pm1.69$ & $0.269^{+0.015}_{-0.014}$ \\ 
BAO-all &$68.11\pm 0.86$ & $0.268^{+0.009}_{-0.010}$\\
\hline
\end{tabular}
\caption{Constraint on $H_0$ and $\Omega_m$ from different BAO datasets. }
\label{tab:bao_all}
\end{table}

\section{Summary and Discussion}

We accumulate all of the BAO data, including 6dFGS, WiggleZ, BOSS DR11 clustering of galaxies and quasar Lyman-$\alpha$ forest lines, SDSS DR7 2dMPS and BOSS DR9 2d2pCF, and find that the Hubble constant is precisely determined by the BAO data alone: $H_0=68.11\pm0.86$ km/s/Mpc, a $1.3\%$ determination. Our result is consistent with Planck \cite{planck} and that given by Efstathiou in \cite{H0_Efstathiou} where the revised NGC 4258 maser distance was used. Our result is also not so different from that obtained by Bennett in \cite{H0_Bennett}. But there is around $2.4\sigma$ tension compared to the Hubble constant given by Riess et al. in \cite{HST_Riess_2011}.

The present matter energy density is also tightly constrained by the BAO datasets alone: $\Omega_m=0.268^{+0.009}_{-0.010}$. This result is nicely consistent with the combination of supernova Union2.1 compilation of 580 SNe (Union 2.1) \cite {union2.1} ($\Omega_m=0.277^{+0.022}_{-0.021}$), but has around $2.6\sigma$ tension with Planck \cite{planck} ($\Omega_m=0.315^{+0.016}_{-0.018}$).

Finally, we want to stress that the high-redshift BAO and the simultaneous measurements of $H(z)$ and $D_A(z)$ from the two-dimensional 2-point correlation function and two-dimensional matter power spectrum are very powerful tools for probing cosmology. We hope that more data about them will be provided in the near future.

\vspace{5mm}
\noindent{\large \bf Acknowledgments}

We would like to thank Ya-Zhou Hu, Miao Li, Sai Wang and Zhen-Hui Zhang for useful conversations. 
This work is supported by the project of Knowledge Innovation Program of Chinese Academy of Science and grants from NSFC (grant NO. 11322545 and 11335012).

\newpage

\end{document}